# Non-volatile superconducting tunnelling magnetoresistance memory enabled by exchange-field gap engineering

Sonam Bhakat[1], Pushpak Banerjee[1], Ahmedullah Aziz[2], Jackson Miller[3] and Avradeep Pal*[1,4]

[1] Department of Metallurgical Engineering and Materials Science, Indian Institute of Technology Bombay, Powai, Mumbai, Maharashtra – 400076, India
[2]Department of Electrical Engineering and Computer Science, University of Tennessee, Knoxville, Tennessee 37996, USA
[3]Paihau-Robinson Research Institute, Victoria University of Wellington, PO Box 33436, Petone 5046, New Zealand
[4] Centre of Excellence in Quantum Information, Computation, Science and Technology, Indian Institute of Technology Bombay, Powai, Mumbai, Maharashtra – 400076, India

**Abstract**
Scalable, low-dissipation memory operating below 4K is a critical requirement for superconducting and quantum computing systems. Existing cryogenic memory technologies rely on CMOS derivatives or hybrid architectures that incur leakage, refresh overhead or limited compatibility with superconducting logic. Here we demonstrate a superconducting tunnelling magnetoresistance device that functions as a non-volatile cryogenic memory element across the full superconducting temperature range. By integrating a de Gennes spin valve with a superconducting tunnel junction in a current-perpendicular-to-plane geometry, we realize exchange-field-controlled modulation of the superconducting energy gap, producing two magnetically switchable gap voltages and robust quasiparticle tunnelling magnetoresistance down to 0.25K. The device operates at millivolt bias with nanowatt-level read power and zero standby dissipation. Its vertical junction architecture and Nb-based materials platform enable compatibility with superconducting logic and scalable cryogenic memory arrays.

**Introduction**
The rapid developments in quantum computing and superconducting digital logic have exposed a critical bottleneck in cryogenic hardware: the absence of scalable, energy-efficient, non-volatile memory technologies compatible with operation below 4K. Existing cryogenic memory approaches[1] rely largely on CMOS derivatives or hybrid architectures that incur static power dissipation, thermal overhead and architectural incompatibility with superconducting circuitry. As superconducting processors approach system-level integration, the development of native cryogenic memory elements has become a pressing technological challenge.

Magnetic tunnel junctions underpin modern non-volatile memory at room temperature through tunnelling magnetoresistance (TMR)[2–4], where relative magnetic alignment controls electronic transport. An analogous device principle within superconducting electronics has remained elusive. While superconducting spintronics enables proximity induced magnetic control of superconducting properties[5–8], and conversely - superconducting control of magnetism[9,10]; demonstrated device implementations of superconducting spin valves have largely been limited to current-in-plane (CIP) geometries[11] in which resistive switching vanishes deep in the superconducting state. As a result, most superconducting spin-valve devices cease to function precisely in the low-temperature regime where cryogenic electronics must operate. This has propelled the need for extensive research into new device concepts, often with greater control over the superconducting transition temperature modulations[12,13], and devices with absolute switching characteristics[14–18]. In addition to the above, CIP devices are inherently incompatible with vertically scalable, high-density memory architectures based on current-perpendicular (CPP) transport.

Here we demonstrate a superconducting TMR device that overcomes these fundamental limitations by exploiting exchange-field control of the superconducting energy gap rather than resistance alone. By integrating a de Gennes spin valve[10,18–22] with a superconducting tunnel junction in a CPP geometry, we convert magnetization-dependent critical temperature modulation into magnetically switchable quasiparticle tunnelling spectra. This architecture yields two distinct superconducting gap voltages corresponding to parallel and antiparallel magnetic configurations, enabling non-volatile quasiparticle tunnelling magnetoresistance (QTMR), starting from the superconducting transition temperature and down to the lowest temperatures. The device operates as a superconducting analogue of a conventional TMR element, but with dissipation-less electrodes and intrinsic compatibility with cryogenic logic. The demonstrated bias dependence of QTMR further enables multi-level state encoding, providing a route toward in-memory and neuromorphic computing architectures at cryogenic temperatures. By establishing exchange-field engineering of the superconducting gap as a device-level control parameter, this work introduces a scalable platform for non-volatile superconducting memory and switching technologies compatible with next-generation cryogenic computing systems.

## Device concept

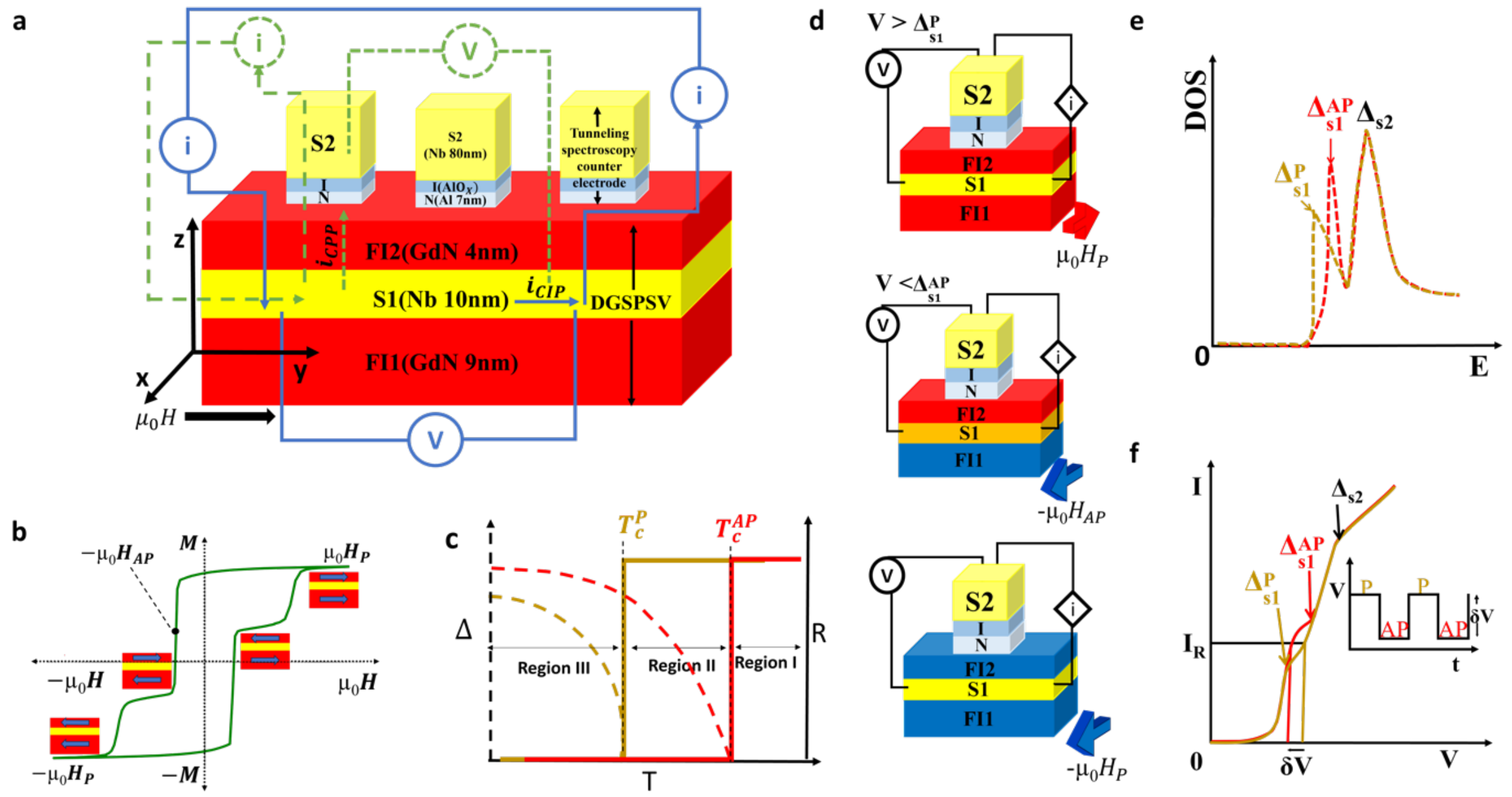

Figure 1|: a) A cartoon representation of a fabricated de Gennes spin valve, with FI1/S/FI2/N/I/S layers for probing both CIP (measured using a current source and voltmeter in blue colour which probes the S1 layer) and CPP characteristics (simplistically demonstrated by using a current source and voltmeter in green colour, which probes the characteristics of the S1/FI2/N/I/S2 tunnel junction. In actual experiment, the differential conductance of the tunnel junction is measured by means of an AC+DC adder box and a lockin amplifier) b) Expected magnetic characteristics of the unpatterned sample. c) Cartoon illustration of CIP resistivity of the S1 layer (Red and golden solid lines), and superconducting gap behaviour of the S1 layer (Red and golden dashed curves); along with various regions of operation of the device d) A cartoon representation of the gap voltage dependence on relative magnetization orientation of the S1 layer. Lighter (darker) shade of yellow in S1 layer at fields corresponding to P (AP) orientations of the FI layers point to a lower (higher) gap voltage of the S1 layer e) Expected relative magnetization orientation dependent gap structure in the CPP measurement geometry f) Expected relative magnetization orientation dependent IV curves in the CPP measurement geometry. Inset shows possible relative magnetization orientation dependent bistable switching behaviour of the device due to gap tunability in S1 layer.

A de Gennes spin valve consists of a superconducting (S) thin film sandwiched between two ferromagnetic insulating (FI) layers[19]. Depending on the relative orientation of the FI layers, the exchange field in the S layer can ideally vary from zero to a maximum value for the anti-parallel (AP) and parallel (P) configurations respectively. Thus far, all known demonstrations of de Gennes Spin valves have been in the CIP geometry[10,18,20–22]. From a transport perspective, the measurable physical quantities in the CIP geometry are resistance (R) and transition temperature ($T_c$). Due to the magnetic orientation dependent variation of exchange field, the $T_c$ of the device is highest and lowest for the AP state ($T_c^{AP}$) and P state ($T_c^{P}$) configurations respectively. Hence, no perceptible resistance difference between the AP and P states appear in these devices below a particular temperature ($T < T_c^{P}$). This presents a major drawback which limits the operation regime of such CIP devices to a narrow range of temperatures ranging between $T_c^{P}$ to $T_c^{AP}$.

However, a superconductor with tuneable $T_c$ should also exhibit a tunable superconducting gap parameter ($\Delta$). Since the superconducting gap and $T_c$ are inextricably linked together by the BCS relation $\Delta_0 = 1.76 k_B T_c$; in a de Gennes spin valve, the superconducting layer will have its $\Delta$ to be dependent on the relative orientation of the FI layers; with $\Delta^{AP}$ ($\Delta^{P}$) corresponding to the higher (lower) $T_c$ states, when the FIs are in AP (P) configurations respectively. Most notably, in the instance the device exhibits a perceptible and significant $\Delta T_c$; starting from the highest temperature of operation ($T_c^{AP}$) and right down to absolute zero, the relation $\Delta^{AP} > \Delta^{P}$ must always hold true. Hence, irrespective of measurement temperature, the gap voltage of S1 layer can be switched by the application of a magnetic field. This change of attaining either the $\Delta^{AP}$ and $\Delta^{P}$value, can be measured by means of tunneling spectroscopy, and used to obtain superconducting TMR and described below.

The above discussion forms the basis of our novel device concept which is illustrated in detail through Figure 1 by means of a series of cartoon depictions. In Figure 1a, we show the typical device geometry which will be used to demonstrate both CIP and CPP behaviours. A multi (six) layered thin film stack (FI1/S/FI2/N/I/S) is grown, and mesa-type tunnel junctions are fabricated and wire bonded as shown in Figure 1a to facilitate the study of the bottom thin-S layer (S1 in Figure 1a) sandwiched between two FI layers having dissimilar switching fields as shown in Figure 1c. Measurement in CIP geometry entails four-probe resistance measurement of the FI1/S1/FI2 de Gennes spin valve. The S1 layer is expected to show two different transition temperatures, as shown in Figure 1c. In the CIP mode, the S1 layer has three regions of operation. In Region I (above $T_c^{AP}$), there is no superconductivity in the S1 layer. In the range $T_c^{P} < T < T_c^{AP}$, marked Region II, the device should switch between the normal and superconducting states depending on relative magnetic orientation of FI layers and demonstrate non-volatile bistable behaviour – as demonstrated in two prior works of ours[18,22]. In Region III, when $T < T_c^{P}$, the S1 layer always remains in the zero-resistance state, and this is immune to any changes in relative magnetic orientations of the FI layers.

In the CPP geometry, the device shown in Figure 1a is essentially a series connection of two superconducting tunnel junctions. The FI1 layer being an insulator, does not participate in the transport process. Starting from the bottom, we have a S1/F2/N junction, which is connected in series with a N/I/S2 junction; where the N layer serves as a common link between the two. Such a device allows tunnelling spectroscopy of both S1 and S2 superconductors. The S1 layer superconducting gap is expected to be a function of relative magnetic orientations of the FI layers, as shown in Figure 1d (lighter shade of yellow in S1 layer is indicative of low gap superconductor $\Delta_{S1}^{P}$), as compared to deeper shade of yellow ($\Delta_{S1}^{AP}$)). In Figure 1e, an indicative cartoon tunneling spectrum of the device is shown, where a clear difference in the spectral features presents as a result of the changing FI layer orientations. In terms of measurable IV characteristics, as shown in Figure 1f, this should translate to an appreciable voltage bulge between voltages corresponding to $\Delta_{S1}^{P}$ and $\Delta_{S1}^{AP}$. This implies that clear bistable switching modes can be achieved in these devices for a certain range of current bias corresponding the region of voltage bulge. Owing to the fact that the relation $\Delta_{S1}^{AP}(T) > \Delta_{S1}^{P}(T)$ always holds true (as shown in Figure 1b); such a switching behavior should persist at all temperatures $T < T_c^{AP}$ (encompassing both Regions II and III). This should make such a device a very attractive candidate for cryogenic memory, cryogenic switching and several other superconducting spintronics applications.

**CIP characteristics of the de Gennes spin valve**

To realize the device concept, we grow a multi-layered stack of GdN (9nm)/ Nb (10nm)/ GdN (4nm)/Al (7nm)-AlOx/ Nb (100nm) on top of a buffer layer of AlN. This stack is then subjected to a multi-step lithography process to fabricate mesa type square shaped tunnel junctions with edge length of $5\mu m$ (Please refer to the methods section for details on growth and device fabrication). The FI2 layer thickness is deliberately kept high (> 3nm) in order to suppress Cooper pair tunneling[23]. In Figure 2, we show the CIP characteristics of the device (where only three functional layers GdN (9nm)/ Nb (10nm)/ GdN (4nm) – come into play), which essentially translates to the magneto-transport study of the bottom S1 layer. Figure 2a shows a colour map of the S1 layer where magnetoresistance has been studied as a function of temperature. The distinctive features of typical CIP de Gennes spin valves are clearly visible. These include a) a clear difference in P and AP state $T_c$ with $T_c^P < T_c^{AP}$ b) a clear broadening in second switching field (referred to as $H_{c2}$) with decreasing temperatures in the range $T_c^P$ to $T_c^{AP}$ , while the first switching field (referred to as $H_{c1}$) stays constant c) a sharp switch to the $T_c^{AP}$ state at $H_{c1}$. In Figure 2b, we compare magneto-resistance and magnetization measurements. The switching fields at various temperatures match well with switching in magnetizations and increase in second switching field (deemed to be that of 4nm GdN from comparisons of magnitude of switchable magnetizations) are almost identical in both measurements. Individual RT measurements indicating $T_c^P$ and $T_c^{AP}$ of the S1 layer are shown in Figure 2c. These features are a reproduction of CIP device characteristics of the same trilayer system presented in a prior publication[22].

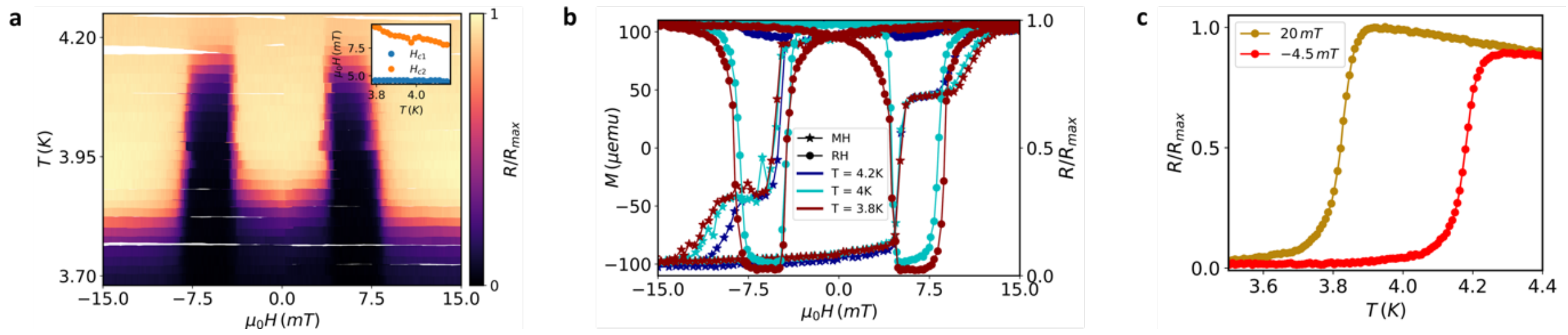

Figure 2: CIP behaviour of the AlN (20nm)/ GdN (9nm)/ Nb (10nm)/ GdN (4nm)/Al (7nm)-AlOx/ Nb (100nm) device a) Colour map shows Resistance vs magnetic field behaviour measured at various temperatures (3.7K to 4.25K) of S1 layer in CIP mode. Measurements are only shown for 0 to +15mT and 0 to -15mT sweeps of field. Inset to Figure a show the temperature dependence of the two switching fields which are extracted from the colour map b) Combined CIP magneto-resistance (patterned sample) and magnetization behaviour (of unpatterned sample) at 3.8K, 4K and 4.2K. c) Two line-cuts from Figure 2a to demonstrate the R vs T measurements at P and AP fields (+15mT and -4.5mT) respectively.

**Demonstration of superconducting gap tuneability in CPP geometry**

In Figure 3a, we show the color map of temperature dependent normalized differential conductance of the device for its P and AP states measured in CPP geometry (refer to Figure 1a). The colour map is shown for voltage range (< 1.6mV) where gap features corresponding to the thin S1 layer is visible. The evolution of two distinctly different gap voltages, one originating at 3.8K ($\Delta^P$) - corresponding to P state (upper panel of Figure 3a), and the other originating at 4.2K ($\Delta^{AP}$) - corresponding to AP state (lower panel of Figure 3a) and saturating to markedly different gap voltage values at the lowest temperatures are evident. To gain further insight on the magnetic tunability of the superconducting gap of the S1 layer, in Figure 3b, we show the differential conductance plotted as a colour map at various external magnetic fields, recorded at a constant temperature (0.26K). At approximately 1.1mV, a visible change from the background can be seen as spike like features at the AP fields on either side (approximately +-5mT), signifying a sudden increase in conductance as compared to the background conductance at all other magnetic fields. This feature occurs due to the sudden switch of the S1 gap voltage from $\Delta^P$ to $\Delta^{AP}$ at the AP fields. In Figure 3c, we show P and AP field line cuts of differential conductance measurements in Figure 3b, at 4K (lower panel) and 0.26K (upper panel). This demonstrates that magnetically tuneable gap voltages can indeed be achieved in our device at all temperatures lower than $T_c^{AP}$. The S2 layer should remain unaffected by changes in relative magnetization orientations of the FI layer. This is observed when the same Figure 3a is shown for larger voltage values and is shown in Supplementary Figure S1. Please refer to Supplementary Figures S2 for dI/dV plots at the P and AP states for various temperatures; and to Supplementary Figure S3 for more elaborate colormaps as those presented in Figure 3b, at several other temperatures.

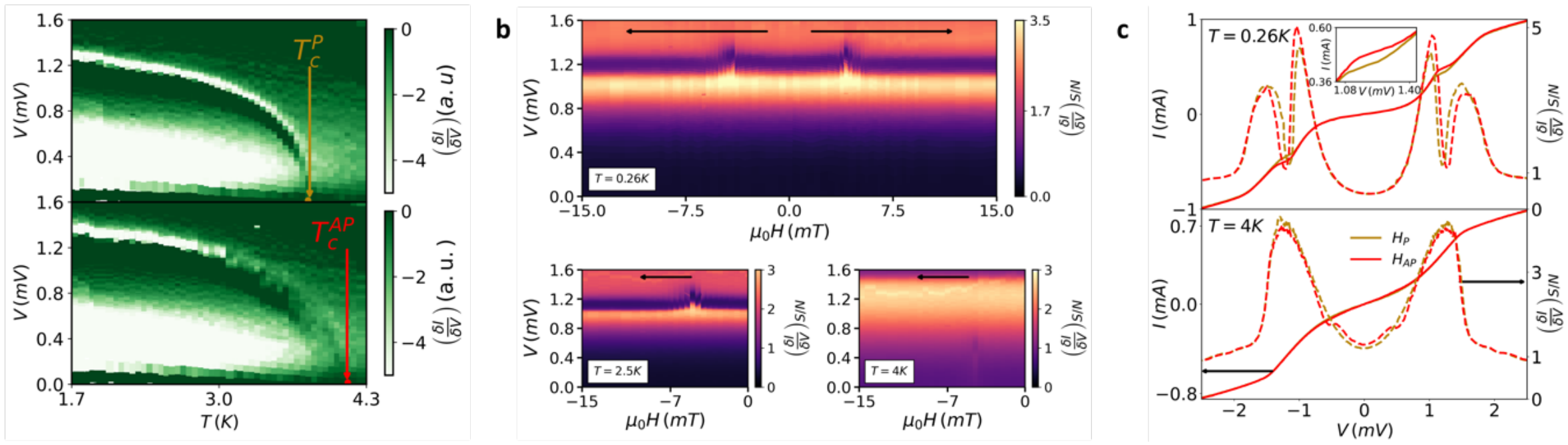

Figure 3: CPP behaviour of the device a) Top (bottom) panel shows colormap of normalized differential conductance measurements done at fields corresponding to the P (AP) orientation of F1 and F2 layers at several temperatures from 1.7K to 4.3K. b) Differential conductance colour maps as a function of bias voltage and in-plane magnetic field at representative temperatures T=0.26K (top), 2.5K (bottom left), and 4K (bottom right), showing pronounced enhancements of normalized differential conductance signals at fields corresponding to AP orientation between FI1 and FI2 layers. Black arrows indicate field sweep direction. c) Current–voltage characteristics (solid lines, corresponding to left y axis) and associated normalized differential conductance spectra (dashed lines, right y axis) for the P state ($H_P$, gold) and AP states ($-H_{AP}$ , red) at T=0.26K (top) and 4K (bottom). Inset in top panel: magnified view of the experimentally measured voltage bulge region showing distinct IV features for P and AP states.

**Bistable switching behaviour of TMR in CPP device**

Switching experiments on de Gennes spin valves have shown evidence of non-volatile bistable characteristics[18,21,22]. However, as previously remarked, due to the CIP nature of these devices, such non-volatile switchable bistable resistance has always been restricted to a narrow range of temperatures $T_c^P < T < T_c^{AP}$. In the case of our device, in the CPP geometry, no temperature restriction of magnetic switching of gap voltage is expected. To test this hypothesis, we apply a constant read current ($I_r$) to our device such that $I(\Delta^P) < I_r < I(\Delta^{AP})$, and measure the zero-field resistance after cycling the magnetic fields between $\pm H_{AP}$ (equivalent to $\pm H_{c1}$), after initially saturating both magnetic layers by applying a magnetic field of $H_P = +15mT$. Here $I(\Delta^P)$, and $I(\Delta^{AP})$ roughly correspond to the current bias required to access the $\Delta^P$ and $\Delta^{AP}$ voltages of the S1 layer. This procedure amounts to measuring the tunability and bistable nature of the quasiparticle tunnelling magneto-resistance (QTMR) of the device. The QTMR is defined as $QTMR\ (\%) = \frac{R_{AP}-R_P}{R_P} \times 100$. As shown in Figures 4a and 4b, the expected bistability of QTMR is measurable down to the lowest temperatures.

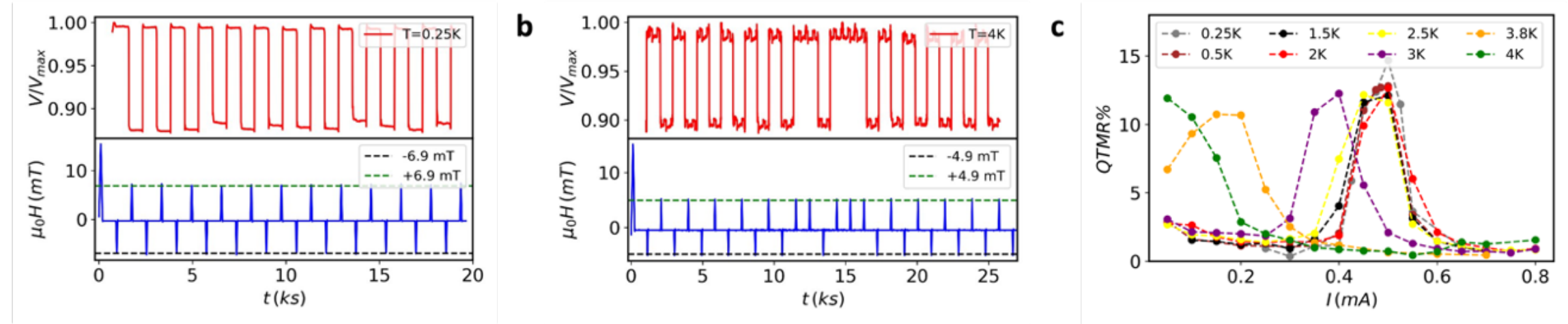

Figure 4| Demonstration of non-volatile bistable switching in CPP mode, at (a) T=0.25K (Region III of Figure 1c) and (b) T=4K (Region II of Figure 1c), under periodic in-plane magnetic field sweeps. Top panels: normalized voltage $\frac{V}{V_{max}}$ showing sharp transitions between P and AP states. Bottom panels: applied magnetic field $\mu_0 H$ versus time, with dashed lines indicating the fields used for cycling for each temperature. c) Bias current dependence of the QTMR ratio, with QTMR (%), measured at different temperatures, showing a pronounced peak that shifts with temperature and decreases in magnitude upon approaching $T_c$.

As evinced through the cartoon IV characteristics shown in Figure 1f, the magnitude of QMTR should be current bias dependent. Moreover, at the highest temperatures of operation $T \sim T_c^{AP}$ , the device should exhibit the lowest magnitudes of $\Delta^{AP}$ and $\Delta^P$, and hence QTMR should appear at low magnitudes of $I_r$. Furthermore, if $\Delta^{AP}$ ($\Delta^{P}$) exhibit typical BCS like behavior, both these magnitudes should saturate at low temperatures. This should be reflected through saturated QTMR at approximately constant $I_r$ below a particular temperature. The QTMR should also only be observable between only a certain range of current biases corresponding to $I(\Delta^{P})$, and $I(\Delta^{AP})$. We show all of these expected device behaviors in Figure 4c for the entire temperature range of device operation. The slight increase in QTMR at the lowest temperatures can be attributed to lowering of thermal noise at lower temperatures.

**Discussion on dI/dV spectral features**

In FI/thin-S structures, due to FI proximity-induced exchange fields, the thin-S1 layer should exhibit quasiparticle spin split DOS[7,24–28]. Such features in DOS can show up in tunnelling spectroscopy measurements, and a full-fledged theoretical fitting of the IV and differential conductance curves should consider these features. However, Nb based superconductors are known to have significantly large spin orbit scattering[29] and Dynes parameter[30] and other factors like finite background DOS[31], due to proximity coupling, which leads to broadening of the DOS features. The combination of all these can contribute to masking or substantial broadening of spin splitting features in the tunnelling conductance spectrum. Hence, when compared to the relatively larger effects of appearance of two distinct magnetically tuneable gap voltages; manifestations of spin degeneracy lifted quasiparticle DOS can be considered second order effects which will not have significant bearing on the main results presented here. Please refer to Supplementary Figures S4 and S5 and discussions therein, for a simplistic model of tunnelling conductance for this device which tries to reproduce the measured spectra for P and AP states.

**Cryogenic memory operation and in-memory computing compatibility**

The device's non-volatile nature of QTMR presents a highly attractive platform for memory technologies operating at cryogenic temperatures[1]. For a given magnetic configuration, the application of an identical bias voltage results in two distinct current levels which can be distinguished using suitable peripheral superconducting circuitry[32], enabling reliable, zero-standby-power readout. Beyond conventional storage, the intrinsic non-volatility of the device naturally fits itself to in-memory computing paradigms at the array level, where computation and storage are co-located. Such architectures significantly mitigate data movement overheads and enhance energy efficiency-an increasingly critical requirement for next-generation AI hardware. Furthermore, the ability to obtain a bias dependent QTMR (as presented in Figure 4c) enables multi-level state encoding, which can be directly mapped onto synaptic weight levels for neuromorphic computing[33]. This multi-state capability allows the emulation of synaptic plasticity with enhanced information density. When integrated with superconducting circuits, this approach combines the ultra-low-power, high-speed advantages of superconducting electronics[34] with the data-locality and parallelism inherent to neuromorphic computing. As a result, the proposed device architecture presents a compelling pathway toward scalable, energy-efficient cryogenic memory and cryogenic neuromorphic systems[35].

**Conclusions**

In conclusion, we have demonstrated a superconducting TMR device that converts exchange-field modulation in a spin valve into magnetically switchable control of the superconducting energy gap. By integrating a de Gennes spin valve with a superconducting tunnel junction in a CPP geometry, we realize non-volatile QTMR that persists across the entire superconducting regime. Although the present devices employ $5\mu m$ mesa, the operating principle is defined entirely by a vertical FI/S/FI/N/I/S multilayer stack analogous to magnetic tunnel junctions. Gap control depends on layer thickness and interfacial coupling rather than lateral area, indicating no intrinsic barrier to sub-micrometre scaling; area reduction will proportionally lower read current and power while increasing integration density without altering the mechanism. Operating at millivolt bias with nanowatt-level read power and zero standby dissipation, this architecture establishes exchange-field-engineered gap control as a scalable route toward non-volatile cryogenic memory and hardware-ready in-memory architectures for hybrid classical–quantum systems.

## Methods

### Growth of multilayers

Multilayers of the type AlN (15nm)/ GdN (9nm)/ Nb (10nm)/ GdN (4nm)/Al (7nm)-AlOx/ Nb (100nm) layer thickness were grown on n-doped Si substrates with a 285nm thermal oxide. Growth was carried out in an ultra-high vacuum custom designed sputtering system with four DC magnetrons and one RF magnetron, with a base pressure below $2 \times 10^{-9}$ mbar. All layers were grown in the same sputtering run without breaking vacuum. The bottom AlN layer act as buffer layer for growth of GdN. AlN was grown using RF sputtering from an Al target using a 56%Ar and 44% N2 reactive gas mixture, and GdN was grown using a 92%Ar and 8% N2 reactive gas mixture from a Gd target. Nb and Al layers were grown using Ar gas plasma from Nb and Al targets respectively. After the growth of the Al layer, the vacuum chamber was filled with high purity oxygen, and subjected to Pressure-time (Pt) product of 15kPa-sec to grow the AlOx layer.

### CPP device fabrication

The deposited stacks were processed using a four-stage optical lithography process to fabricate square shaped mesa junctions[36]. First, a narrow strip containing all layers was etched out using $CF_4$ for the Nb layers, and Argon ion milling for the Al-AlOx, and GdN layers. In the second step, mesas were defined within this narrow strip, and the top Nb and Al-AlOx layers were etched out. Next, a via was fabricated on top of the mesas, and all mesas were electrical isolated from each other, using $SiO_2$ deposition and lift-off. Finally, a layer of superconducting Nb was deposited and lifted off to contact the mesas through the via defined in the last step.

### Electronic and magnetic measurements

Low-temperature electrical measurements were done in Oxford Teslatron PT cryostat, where a base temperature of 0.26K can be achieved using a He3 insert. Measurements down to 1.7K was carried out in a separate probe, while all measurements below 1.7K was carried out in the He3 insert. For CIP measurements, a current of $100\mu A$ was used. For dI/dV measurements in the CPP geometry, an operational amplifier based summing amplifier was used to enable addition of a constant AC and varying DC bias, which was fed into the device. dV/dI measurements were obtained directly using a lockin amplifier and its reciprocal was numerically calculated to get the dI/dV signal. Magnetic fields were applied parallel to the plane of the films. Switching measurements were performed by using a constant DC read current, and monitoring the voltage output of the device using a voltmeter. Magnetometry measurements presented were performed using the reciprocating sample option (RSO) mode of a commercially available Magnetic Property Measurement System (MPMS) from Quantum Design. The applied magnetic field was isothermally swept from a large positive field (ensuring the GdN present in the sample was saturated) to the same magnitude field in the opposite direction and then returned to the initial condition. Each data point is the average of three separate measurements in which the magnetisation is inferred from a comparison with an ideal dipole response.

### Acknowledgements

AP, SB and PB was financially supported by a Core Research Grant from the Department of Science and Technology, Science and Engineering Research Board, India (Grant No. CRG/2019/004758). JM was supported by the New Zealand Endeavour fund (Grant No. RTVU1810) and SMART Idea Grant (Grant No. RTVU2504), both funded by the New Zealand Ministry of Business Innovation and Employment.

### Author contributions

AP conceived the idea. SB grew the multilayers, fabricated the devices and performed all electrical measurements. PB helped in differential conductance measurements and simulation of the results of tunnelling spectroscopy. JM performed magnetic measurements. AZ helped in identifying practical use cases of the devices. All authors discussed the results and contributed to preparing the manuscript.

Supplementary information to

# Non-volatile superconducting tunnelling magnetoresistance memory enabled by exchange-field gap engineering

Sonam Bhakat[1], Pushpak Banerjee[1], Ahmedullah Aziz[2], Jackson Miller[3] and Avradeep Pal*[1,4]

[1] Department of Metallurgical Engineering and Materials Science, Indian Institute of Technology Bombay, Powai, Mumbai, Maharashtra – 400076, India
[2]Department of Electrical Engineering and Computer Science, University of Tennessee, Knoxville, Tennessee 37996, USA
[3]Paihau - Robinson Research Institute, Victoria University of Wellington, PO Box 33436, Petone 5046, New Zealand
[4] Centre of Excellence in Quantum Information, Computation, Science and Technology, Indian Institute of Technology Bombay, Powai, Mumbai, Maharashtra – 400076, India

## 1. $dI/dV$ spectra showing the conductance of both S1 and S2 superconducting electrodes

Supplementary Figure 1 shows the temperature–dependent differential-conductance spectra for the P (top) and AP (bottom) magnetic configurations. The lower-voltage conductance feature corresponds to the S1 electrode (Figure 3a in main manuscript). The higher-bias conductance feature (starting from approximately 2.35mV) is associated with the thick S2 electrode, remains essentially unchanged over the entire temperature range shown, with no observable dependence on magnetic configuration. This confirms that S2 is magnetically decoupled from the spin-valve structure and serves as a stable reference superconductor, while all magnetic-state–dependent modifications arise exclusively from S1.

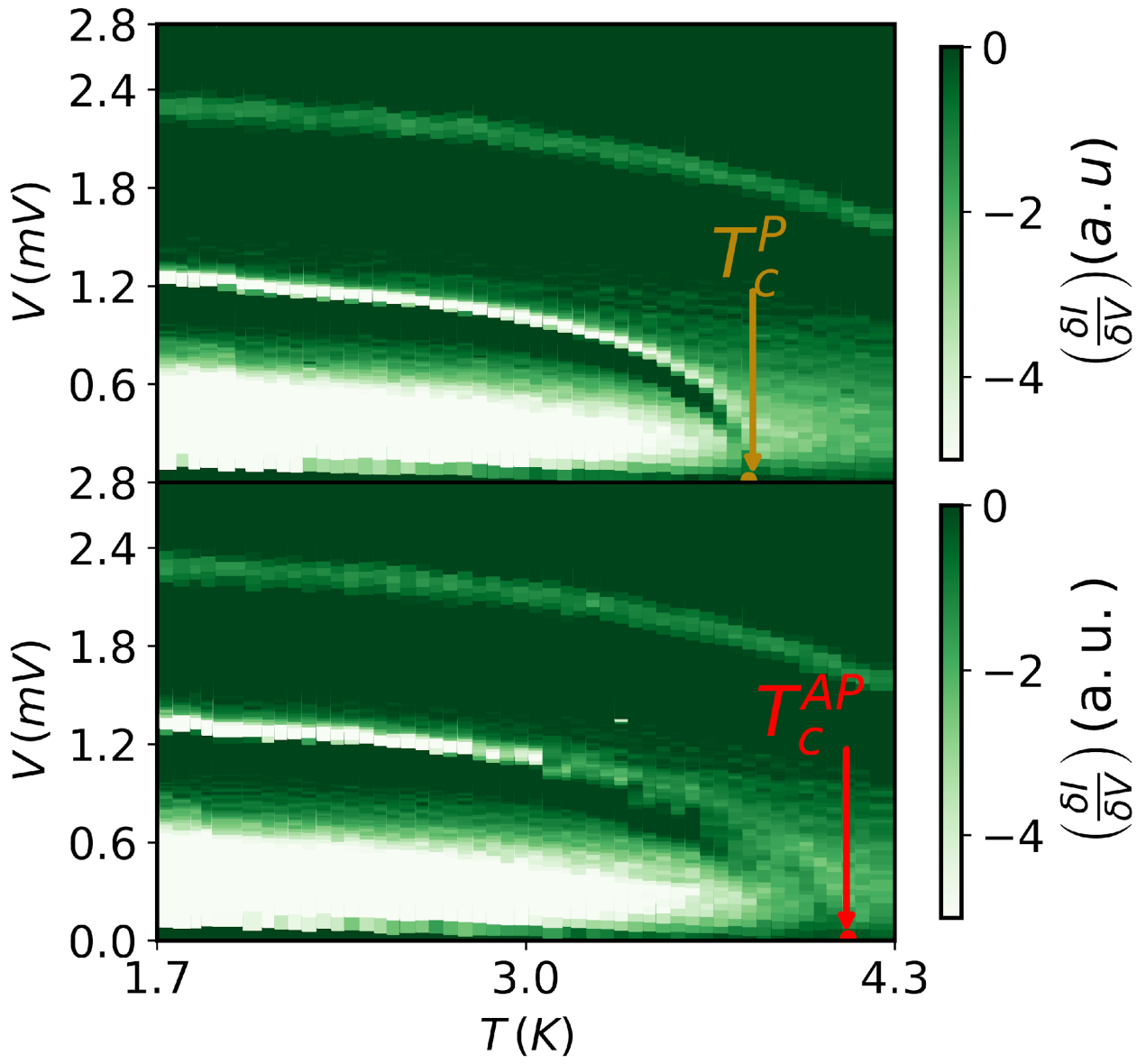

Supplementary Figure S1: Top (bottom) panel shows colormap of differential conductance measurements done at fields corresponding to the P (AP) orientation of F1 and F2 layers at several temperatures from 1.7K to 4.3K.

## 2. $dI/dV$ Spectra at different temperatures at $H_{AP}$ and $H_P$

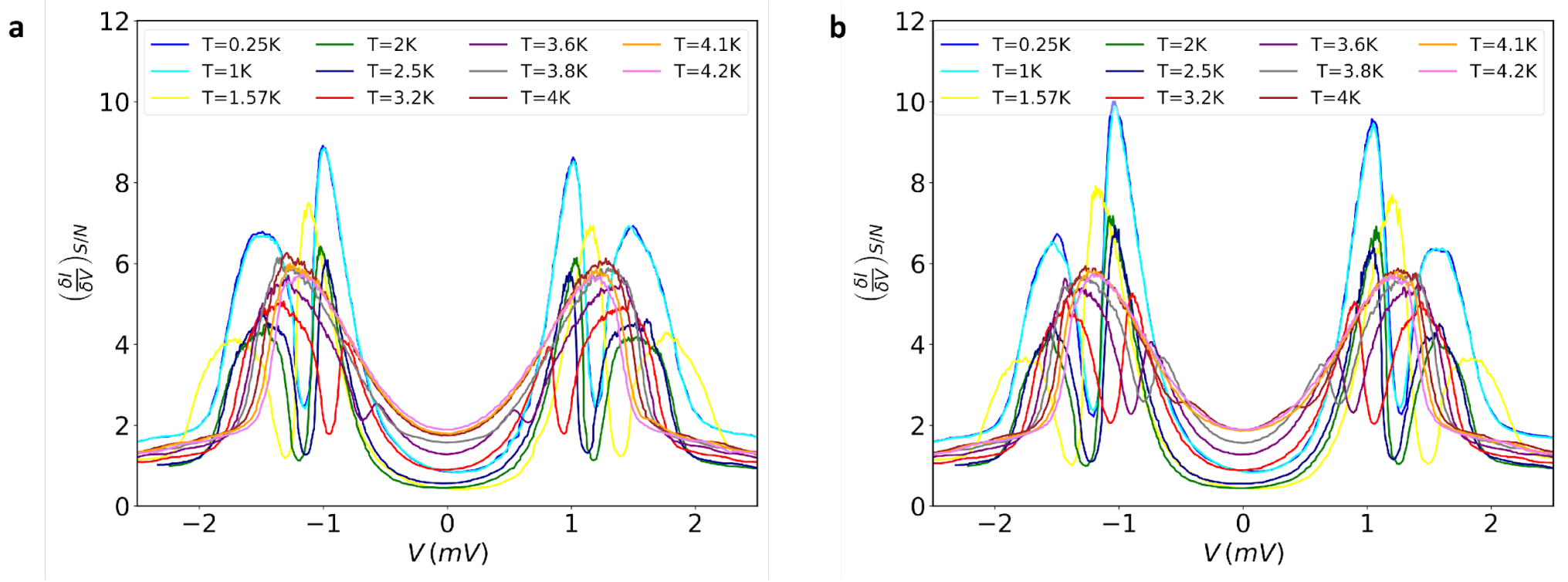

Supplementary Figure S2| Evolution of normalized differential conductance spectra of the device with temperature at two different magnetic fields magnetic fields corresponding to (a) P state (measured at $H_P$) (b)AP state (measured at $-H_{AP}$ after application of $+H_P$).

## 3. $dI/dV$ colormaps of the device under magnetic-field sweeps at various temperatures

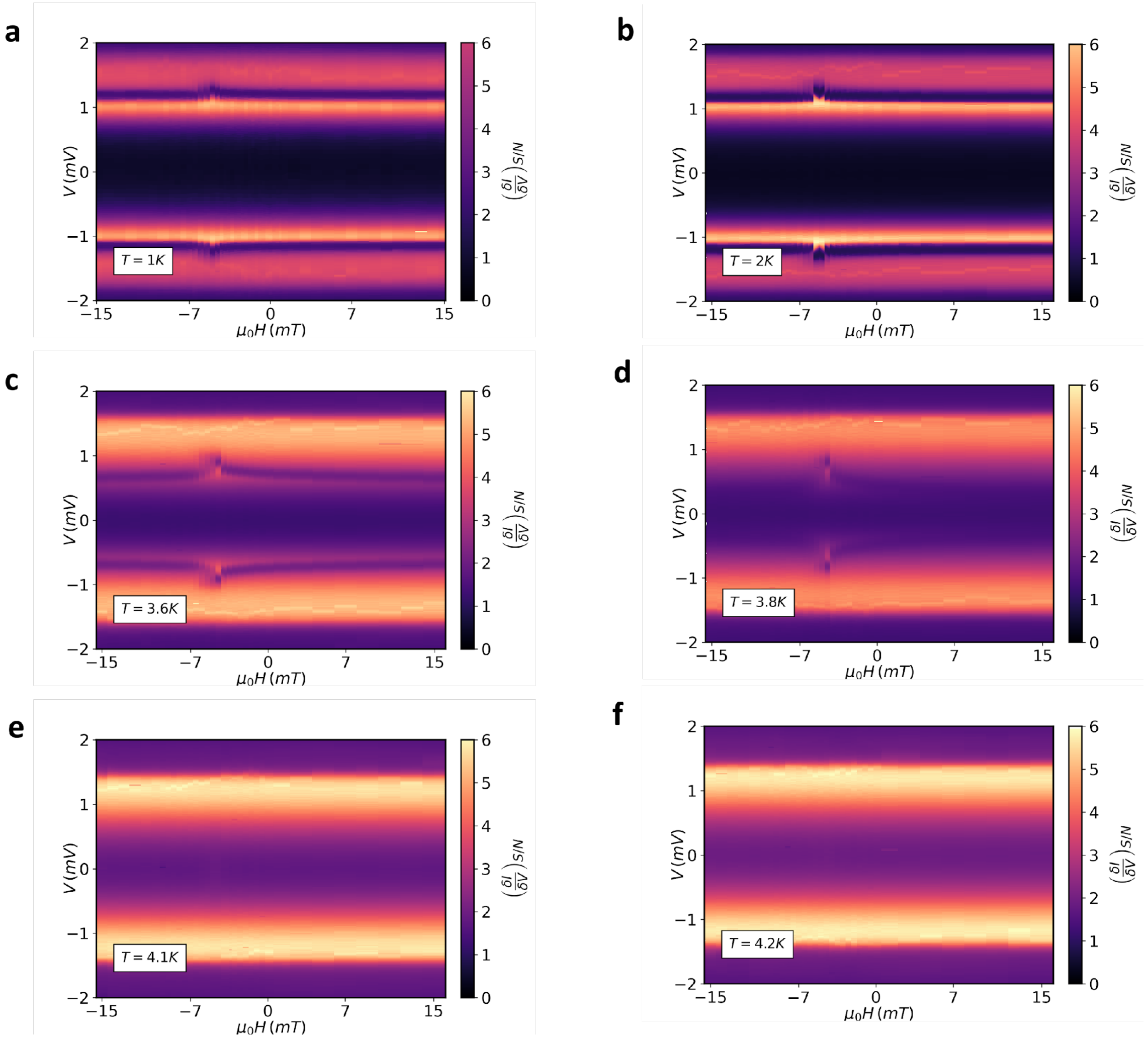

Supplementary Figure S3 | $dI/dV$ colormaps of the device under magnetic-field sweeps, plotted as a function of bias voltage $V$ and in-plane magnetic field $\mu_0 H$ for temperatures ranging from $T = 1K$ to $T = 4.2K$. At low temperatures, the characteristic superconducting coherence

peaks of S1 generate bright, field-dependent ridges that sharpen around the switching fields of the magnetic layers. With increasing temperature, these features progressively weaken and shift toward lower voltages, reflecting the reduction of the superconducting gap. By $T \approx 4.2\ K$, the spectra become nearly field-independent as S1 approaches the fully normal state (close to $T_c^{AP}$).

Supplementary Figure S3 presents the evolution of the differential-conductance spectrum of the device, as a function of magnetic field and temperature. At low temperatures, distinct coherence peaks appear at symmetric (positive and negative) bias voltages, as the FI layers switch magnetization, revealing the magnetic sensitivity of the S1 superconducting density of states. As the temperature increases, the superconducting gap gradually decreases, causing the high-contrast coherence-peak ridges to shift inward and diminish in amplitude. Above approximately 3.8–4.2 $K$, the gap becomes strongly suppressed and the spectra flatten, showing minimal dependence on magnetic field. This temperature-driven collapse of field-sensitive spectral features confirms that the observed modulation originates from superconductivity in the S1 electrode rather than from changes in junction resistance or magnetic background.

4. **Simulated tunnelling spectra of the device**

For a comparative simulation of the observed differential conductance profile with respect to voltage V of the FI1-S1-FI2-N-I-S2 device it was represented with a model where the full device is treated as a combination of two segments:

Part1- FI1-S1-FI2-N with the proximitized superconductor S1, and

Part2- N-I-S2 with the counter superconducting electrode S2.

Thus, the estimated total normalised conductance $g(V)$ is a sum of individual normalised conductance profiles for part1 and part2 which are, $g_1(V)$ & $g_2(V)$ respectively.

$$g(V) = g1(V) + g2(V) \qquad (1)$$

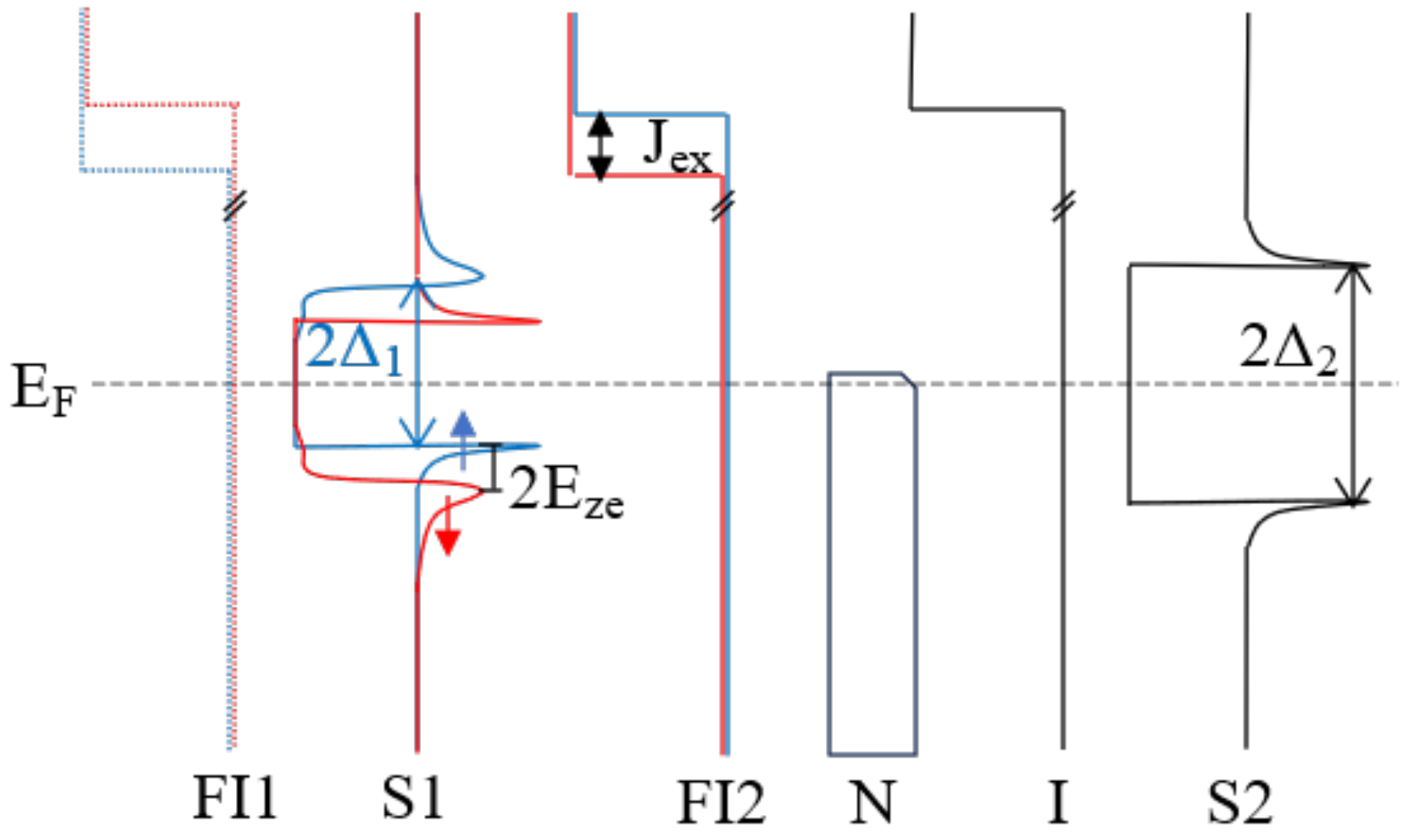

Supplementary Figure S4| A representation of the state spectrum of the FI1-S1-FI2-N-I-S2 structure at equilibrium. Under biased conditions the quasiparticles from the Zeeman split ($E_{ze}$) superconductor S1 (with gap $\Delta_1$) tunnel through ferromagnetic insulator FI2 to normal metal interlayer N; and, again undergo tunnelling through an insulating barrier I before being collected on the superconducting electrode S2 (with gap $\Delta_2$).

Since, both segments are effectively S/I/N type hence following the usual definition [1], the normalized differential conductance for the Part1 is a convolution of the density of states (DOS)- $N1(E)$ of S1 with the partial derivative of the fermi distribution $f$ at temperature T. Where, quasiparticle energy E is with respect to fermi level. However, in Part2 with the superconductor S2 (DOS: $N2(E)$) a provision was made to account for a background conductance using the parameter C [2].

$$g1(V) = N1(E) \star \frac{\partial f(\frac{E+eV}{K_B T})}{\partial V} \quad (2)$$

$$g2(V) = \frac{C + N2(E) \star \frac{\partial f(\frac{E+eV}{K_B T})}{\partial V}}{C+1} \quad (3)$$

In part1, the tunnelling current is considered to be 95% (P) spin polarized by FI2 barrier. A host of inelastic scattering effects within S1 is summed conveniently with a dynes factor of $\Gamma_1$ = 0.01 $\Delta_1$ [3].

$$N1 = \frac{(1-P)N_{1\uparrow} + (1+P)N_{1\downarrow}}{2} \quad (4)$$

$$N_{1\uparrow\downarrow} = Re\left(\frac{u_{\pm} - i\,\Gamma 1}{\sqrt{(u_{\pm} - i\,\Gamma 1)^2 - 1}}\right) \quad (5)$$

The S1 is spin split with a Zeeman energy corresponding to $\mu_B B$ = 0.2 $\Delta_1$. A finite contribution of Spin orbit scattering $b_{so}$ ~ 0.1 $\Delta_1$ is also accounted for [4].

$$\frac{E \mp \mu_B B}{\Delta_1} = u_{\pm} + bso \frac{u_{\pm} - u_{\mp}}{\sqrt{1 - u_{\mp}^2}} \quad (6)$$

Part2 is akin to a usual S/I/N type with equal contribution from both spin channels, however a dynes correction of $\Gamma_2$ = 0.05$\Delta_2$ is also incorporated to address the washed-out peaks.

$$N2 = \frac{N_{2\uparrow} + N_{2\downarrow}}{2} \quad (7)$$

$$N_{2\uparrow\downarrow} = Re\left(\frac{v_{\pm} - i\,\Gamma 2}{\sqrt{(v_{\pm} - i\,\Gamma 2)^2 - 1}}\right) \quad (8)$$

In the thicker electrode S2 a depairing parameter is considered of the order $\zeta$ = 0.01 $\Delta_2$ [4].

$$\frac{E}{\Delta_2} - v_{\pm}\left(1 - \frac{\zeta}{\sqrt{1 - v_{\pm}^2}}\right) = 0 \quad (9)$$

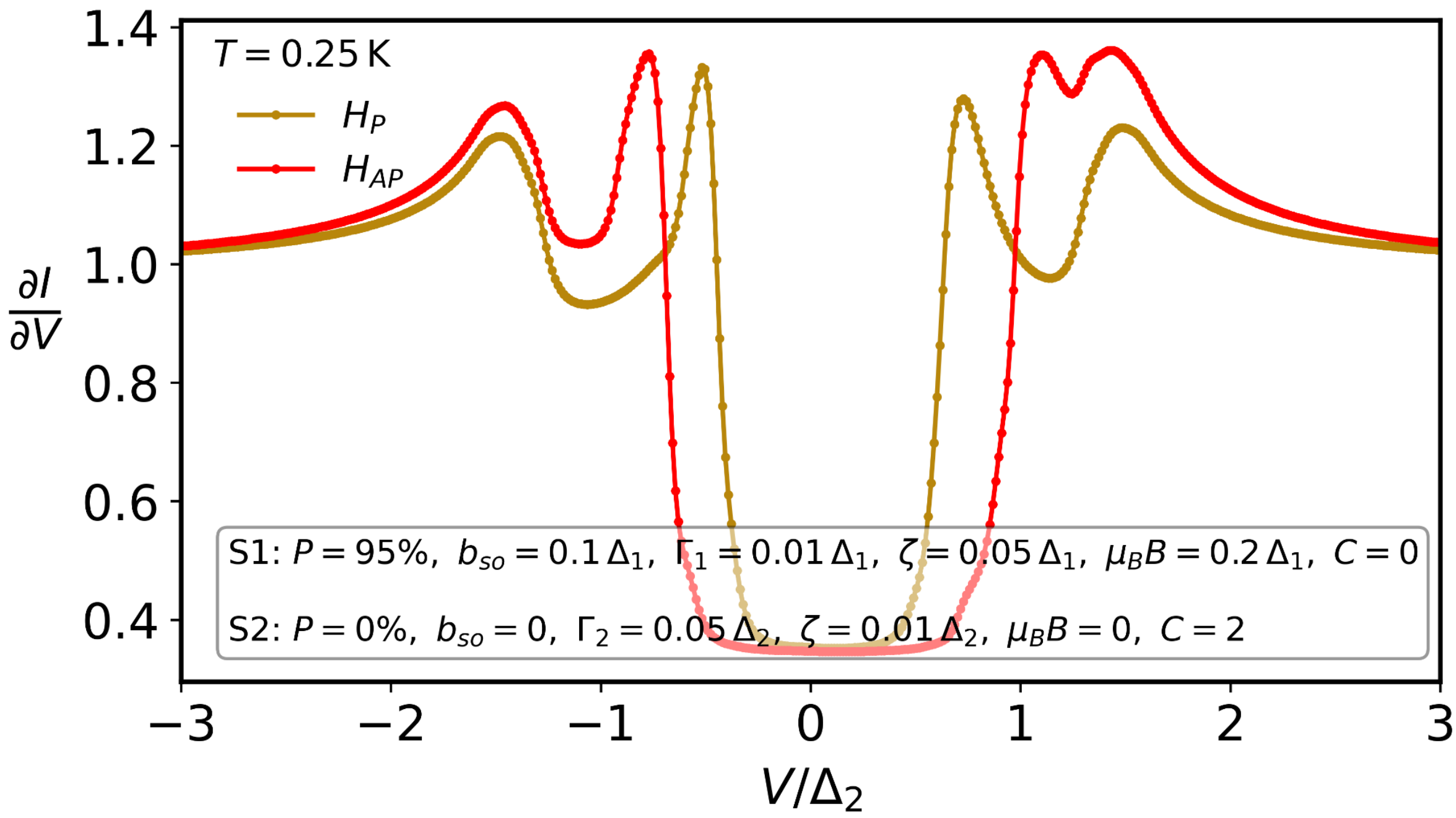

Supplementary Figure S5| Computed differential conductance curves using an effective tunnelling model as discussed above. The brown curve corresponds to the saturated magnetic state $H_P$, while the red curve represents the conductance at $H_{AP}$, the difference between them being the assumed exchange fields in the S1 layer.

**Supplementary References:**